\documentclass[a4paper]{article}

\usepackage{INTERSPEECH2020}
\usepackage{multirow}
\usepackage{graphicx}

\title{Uniphore's submission to Fearless Steps Challenge Phase-2}
\name{Karthik Pandia D S, Cosimo Spera}
\address{Uniphore Software Systems, Chennai, India}
\email{karthikpandia@uniphore.com, cosimo@uniphore.com}

\begin{document}

\maketitle
\begin{abstract}
We propose supervised systems for speech activity detection (SAD) and speaker identification (SID) tasks in Fearless Steps Challenge Phase-2. The proposed systems for both the tasks share a common convolutional neural network (CNN) architecture. Mel spectrogram is used as features. For speech activity detection, the spectrogram is divided into smaller overlapping chunks. The network is trained to recognize the chunks.
The network architecture and the training steps used for the SID task are similar to that of the SAD task, except that longer spectrogram chunks are used. We propose a two-level identification method for SID task. 
First, for each chunk, a set of speakers is hypothesized based on the neural network posterior probabilities. 
Finally, the speaker identity of the utterance is identified using the chunk-level hypotheses by applying a voting rule.
On SAD task, a detection cost function score of $5.96\%$, and $5.33\%$ are obtained on dev and eval sets, respectively. A top 5 retrieval accuracy of $82.07\%$ and $82.42\%$ are obtained on the dev and eval sets for SID task. A brief analysis is made on the results to provide insights into the miss-classified cases in both the tasks.
\end{abstract}


\section{Introduction}

Algorithms for speech processing usually degrades when applied to speech signals that have high noise levels \cite{gong1995speech, zhao2014robust, ramirez2007voice}. Apollo-11 Corpus \cite{sangwan2013houston} is one such data with a huge amount of audio files at various noise levels. Fearless steps data is collected by digitising Apollo mission audio data. A 100 hours of the corpus was released as a part of fearless challenge phase 1 \cite{Hansen2019}. The challenge was conducted to build systems that can work on challenging data. The second phase of the challenge has 4 tasks, namely, speech activity detection, speaker identification, speaker diarization, and automatic speech recognition. This paper describes systems submitted on two tasks - speech activity detection and speaker identification.

Speech (voice) activity detection (SAD) is used as a pre-processing block in almost every speech applications. Various tasks that use SAD as a pre-processor include automatic speech recognition, speaker diarization, speaker recognition, speech enhancement, speech encoding, to mention a few. SAD on noisy speech is still a challenging problem. Different features used in SAD task includes zero-crossing rate \cite{rabiner1975algorithm}, spectral flatness \cite{sadjadi2013unsupervised}, auto-correlation function \cite{kristjansson2005voicing}, short-time average magnitude difference function \cite{orlandi2003maximum},  MFCC, pitch \cite{nelson1995pitch}. A detailed survey of different features used in the SAD task can be accessed in \cite{graf2015features}. On the classification side, GMM \cite{ng2012developing} and artificial neural network (ANN) \cite{ryant2013speech} are the two common approaches. Various ANN-based approaches include convolutional neural network (CNN) \cite{thomas2014analyzing}, recurrent neural network (RNN) \cite{hughes2013recurrent}. Classifiers either work at the frame-level or at the segment level or a frame-level decision followed by post-processing to obtain speech segments. 
The baseline provided for SAD task uses a combo feature \cite{sadjadi2013unsupervised}. The features include time-domain features harmonicity, clarity, and prediction gain; and frequency domain features periodicity and perceptual spectral flux.  This 5-dimensional feature vector is projected to 1-dimensional feature vectors using principal component analysis. The speech and non-speech regions are modelled using a two mixture GMM, one for each class.

Conventional speaker identification techniques are made up of three blocks. First is the feature representation, front-end classification comes next, and finally, back-end classification. MFCC and PLP are the most common feature types used for representation. The front-end classification approaches include UBM-GMM \cite{reynolds2000speaker}, i-vector \cite{kenny2015vector} and SVM-based \cite{campbell2006support}. Score normalization, PLDA \cite{kenny2013plda} are commonly used as back-end processing steps. Most recent approaches are neural network-based approaches \cite{snyder2018x,wan2018generalized}. 
Convolutional neural networks have been used in several works for speaker recognition tasks \cite{lukic2016speaker, salehghaffari2018speaker, dhakal2019near}.
Spectrogram, filterbank energy, and MFCC are common features. 
Despite a lot of advancement in SID problem, it is still a challenging task when the duration of test utterance is short and if the SNR value of the signal is low.  It has been shown that that the detection performance decreases significantly when the duration of the test utterance decreases from 30s to 20s or from 20s and 10s \cite{Sadjadi2019}.
The baseline results provided  by the organizers for the SID task is a CNN-based approach  called \textit{SincNet} \cite{ravanelli2018speaker}. In addition to the usual CNN architecture, SincNet attempts to learn the filterbank using a convolutional layer.

This work explores the effectiveness of CNN for the SAD and the SID task on Fealress steps dataset. In addition to the exploration of CNN architecture, a scoring method is proposed for the SID task.
The rest of the paper is organized as follows. The statistics of the dataset is briefly shown in Section~\ref{sec:data}. The architecture details and the methodology proposed for SAD and SID tasks are given in Section~\ref{sec:prop}. Details about the experiments and the evaluation setup is given Section~\ref{sec:exp}. The results are discussed and analyzed in Section~\ref{sec:results}. Conclusions are made in Section~\ref{sec:conc}.

\section{Dataset description}
\label{sec:data}

\begin{table}[h]
\caption{Description of the dataset used for SAD and SID tasks}
\begin{tabular}{|c|c|c|c||c|c|c|} \hline
&\multicolumn{3}{c||}{SAD} & \multicolumn{3}{c|}{SID} \\ \cline{2-7}
& Train & Dev & Eval & Train & Dev & Eval \\ \hline
\# files & 120 & 20 & 40 & 27336 & 6373 & 8466 \\ 
Duration (h) & 63.56 & 15.2 & 20.4 & 30.62 & 7.15 & 9.5 \\ \hline
\end{tabular}
\label{tab:data}
\end{table}

\begin{figure*}[t]
    \centering
    \includegraphics[width=2.0\columnwidth]{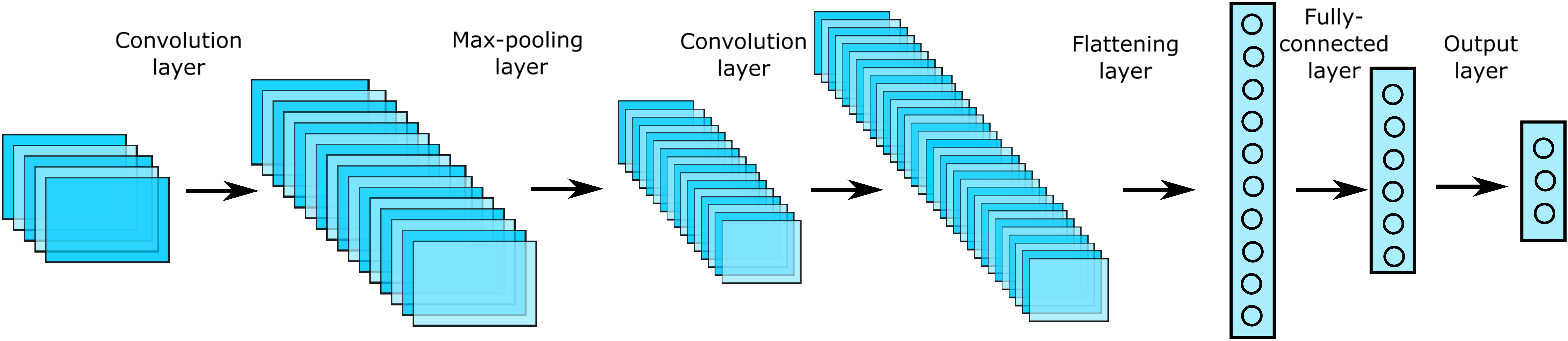}
    \caption{Architecture of the convolutional neural network used for SAD and SID tasks}
    \label{fig:cnn}
\end{figure*}

Table~\ref{tab:data} shows the statistics of the dataset for SAD and SID tasks in Fearless challenge 2. Typically, the duration of the data in each set of SAD is twice as that of the SID task. The average duration per file for SID task is approximately $4s$ for all three sets. The total number of speakers in the dataset is 218. The average per speaker duration for training is $505.6s$, and the standard deviation is $129.5s$. The duration statistics shows that the dataset is 1. Heavily imbalanced with respect to the amount of training data per speaker 2. Challenging in terms of the duration of test utterances.

\section{Proposed Approaches}
\label{sec:prop}

The proposed systems use a convolutional neural network-based classifier. The usage of CNN for speech tasks has been demonstrated long back \cite{lecun1995convolutional}. Though there have been attempts to use CNN in different tasks, the system has to be hand-crafted to achieve the best output. Here we explore the parameters for feature extraction and the network architecture for the challenge data.
The architecture of the network is shown in Section~\ref{ssec:arc}. The speech activity detection and speaker identification systems are described in Section~\ref{ssec:sad} and Section~\ref{ssec:sid}, respectively.

\subsection{Network architecture}
\label{ssec:arc}

The approaches proposed for both speech activity detection and speaker identification tasks use a similar CNN architecture. Several architectures have been tried out on the development data. The architecture that gave the best result is shown in Figure~\ref{fig:cnn}. The architecture consists of $6$ layers. The input is fed to a convolution layer followed by a max-pooling later and another convolutional layer.
The output from the convolutional layer is flattened in the next layer. Then follows a  fully connected layer, which is finally connected to the output layer. The network is trained using the Adam optimizer and cross-entropy loss function.

\subsection{Speech activity detection task}
\label{ssec:sad}

\begin{figure}[h]
    \centering
    \includegraphics[width=\columnwidth]{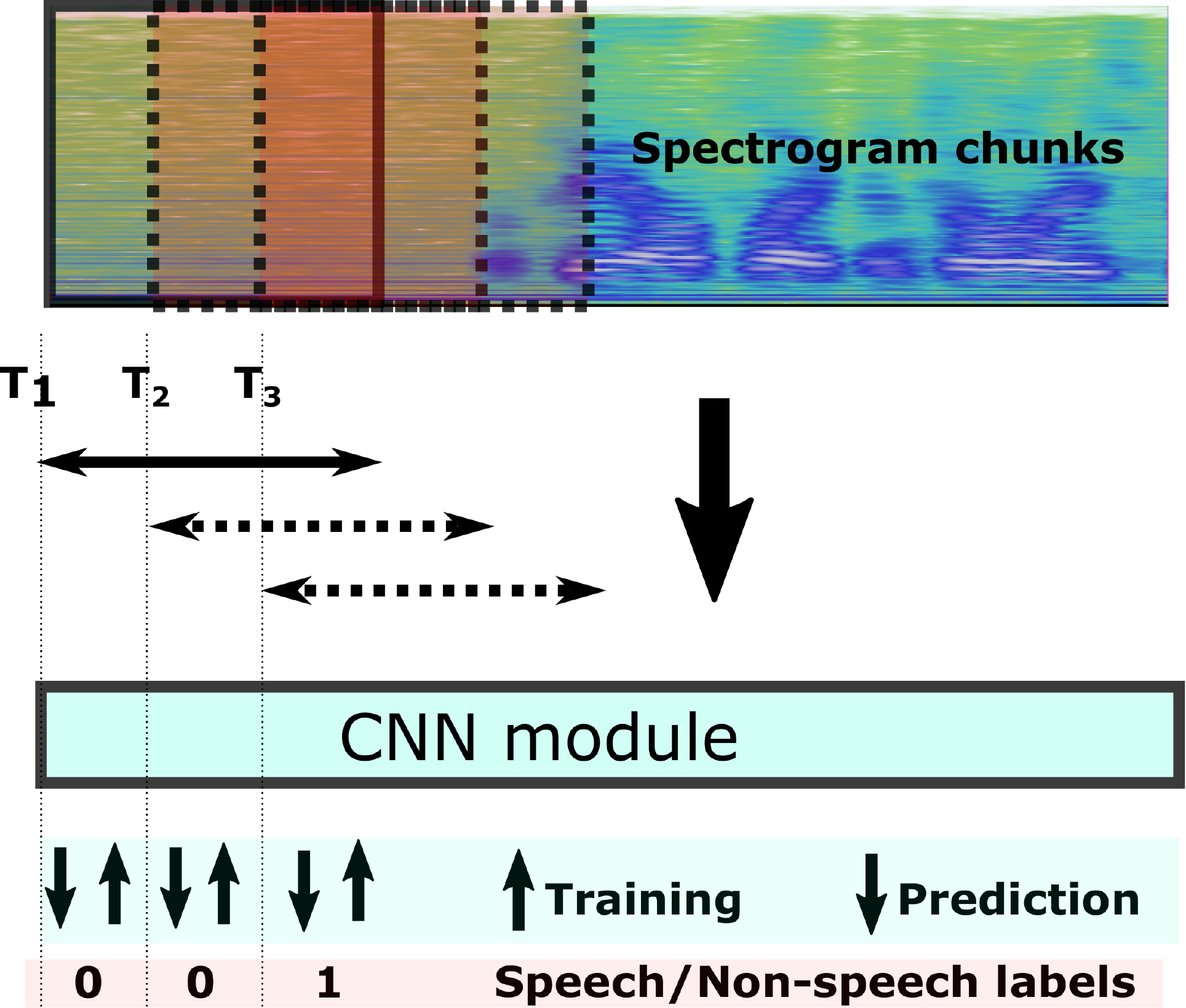}
    \caption{Training and recognition stages for SAD system}
    \label{fig:train}
\end{figure}

On the speech activity detection task, the network is learnt to discriminate speech and non-speech regions.  The flow of the proposed VAD system is illustrated in Figure~\ref{fig:train}. The spectrogram is computed for the entire long audio using a frame size of $50ms$ with an overlap of $10ms$. The long spectrogram is chunked into overlapping spectrograms. The optimal spectrogram chunk size and chunk shift are found to be $320ms$ and $160ms$, respectively.  For each $160ms$ chunk shift, the corresponding label is found by mapping the shift time stamps to that of the ground-truth labels. The network is learnt to classify the chunks of size $320ms$ as either speech $(0)$ or non-speech $(1)$ regions. A subset of the training data is used as a validation set to avoid overfitting of the network. While testing, similar to the training phase, overlapping spectrogram chunks are obtained. These chunks are fed to the network to identify whether the chunk corresponds to speech or non-speech regions. Since the chunks are overlapping; decisions are made for each non-overlapping regions between each pair of consecutive chunks. The training and prediction stages share a common block except for the change in the flow of the labels as shown in the figure.

\subsection{Speaker identification task}
\label{ssec:sid}

\begin{figure}[t]
    \centering
    \includegraphics[width=\columnwidth]{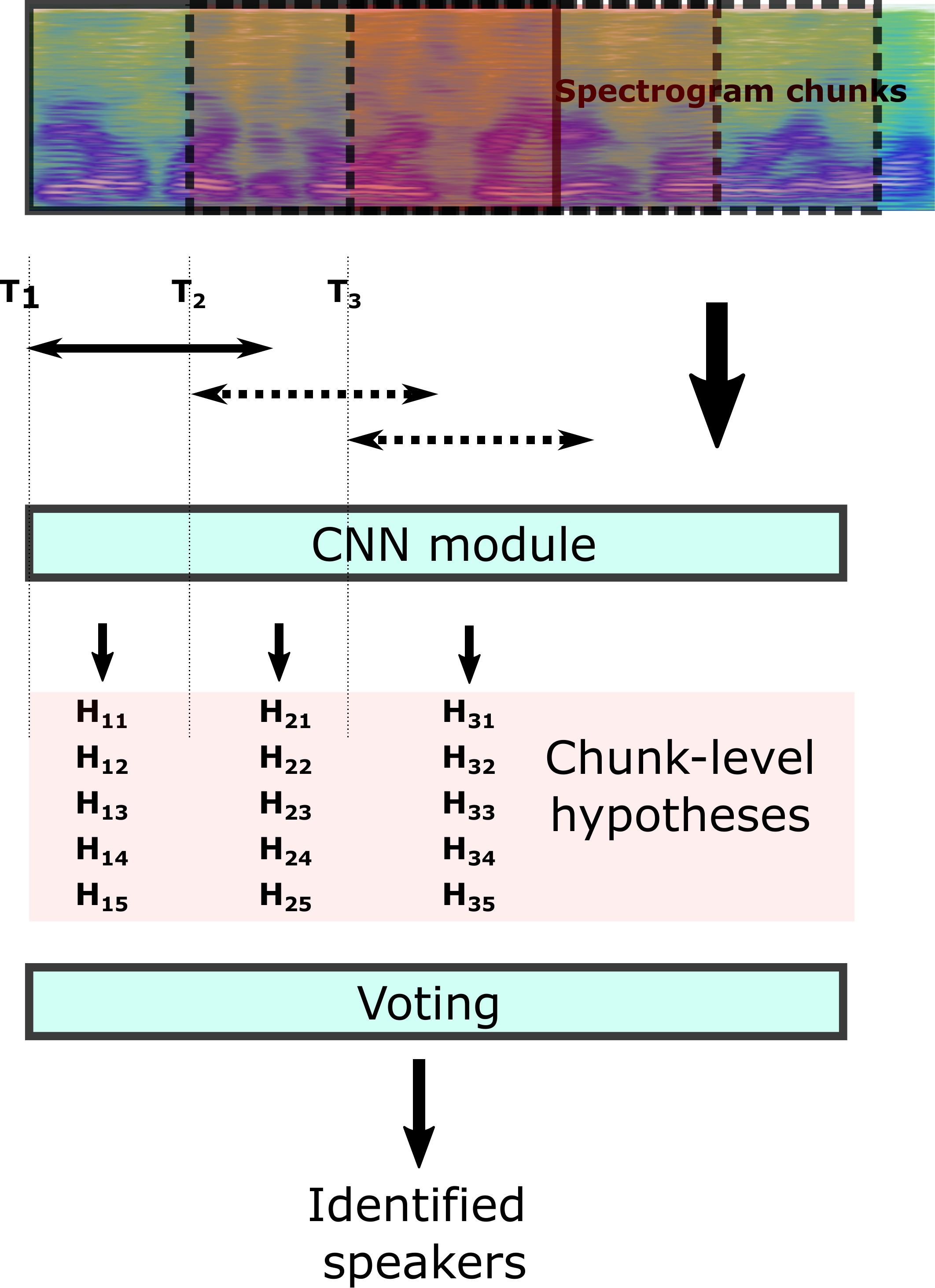}
    \caption{Identification method for the speaker identification task.}
    \label{fig:sid}
\end{figure}

As mentioned in Section~\ref{ssec:arc}, the basic architecture of the network for the SID system is the same as that of the SAD system. The SID systems differ from SAD systems in two aspects. The first difference is in the parameters of Mel spectrogram. The second difference is in the identification stage. The block diagram of the proposed SID system is shown in Figure~\ref{fig:sid}

On the difference in the frame, the frame size used to compute the spectrogram is $25ms$, which is shorter than that of the SAD task. The duration of the spectrogram chunks for SID task is much longer $(1.28s)$ than that of the SAD task. Similarly, the spectrogram shift is higher $(160ms)$ than that of the SAD task.

The training of the network for the SID task is similar to that of the VAD task. The difference is only in the number of output nodes in the output layer, which is equal to the number of speakers in the training data. During the identification stage, unlike a single decision making as in the case of the traditional SID systems, two-level decisions are made. In the first level, the decisions are made at the chunk level for each $160ms$ of speech. The speakers with top 5 likelihood are hypothesized as candidate speakers. The candidates are shown as $H_{ti}$ in the figure, where $t$ and $i$ denote chunk ID and rank of the retrieval, respectively. Hence, for the entire test utterance, a bag of speakers along with the frequency of occurrence is obtained. In the next level, a voting rule is used to decide the top-5 speakers for that utterance.

\section{Experimental setup}
\label{sec:exp}

For the SAD task, the train set provided by the organizers had $120$ files with a total duration of $50$ hours. Out of the $120$ files, the first 100 files are used to train the network, and the rest $20$ files are used to validate the parameters. The tuning of the parameters outside the neural network is performed on the development data which had $30$ files. 
The train and dev sets provided by the organizers are used as train and validation sets of for the SID task.
Tensorflow \cite{tensorflow2015-whitepaper} is used to build CNN.
The best performing system had $32$ filters in both the convolutional layer. Masks of dimension $5\times5$ and $3\times3$ are used in convolutional layer and max-pooling layers respectively. The pen-ultimate layer had 64 nodes for SAD and $500$ nodes for $SID$ tasks, respectively.

The SAD system is evaluated using a detection cost function (DCF) measure. DCF is a weighted combination of false-positive rate and false-negative rate. For a threshold $\theta$, DCF is defined as 
\begin{equation*}
    DCF(\theta) = 0.75 \times P_{FN} (\theta) + 0.25 \times P_{FP} (\theta)
\end{equation*}
where, $0.75$ and $0.25$ are the weights of the respective error rates.
A tolerance collar of $250ms$ is used to compute the DCF.

Top-5 detection accuracy is used as the evaluation measure for SID task. The formula of the measure is given below.
\begin{equation*}
    Accuracy = \frac{\sum_{i \in S}^{} N_{ref} (i)}{\sum_{i=1}^{M} N_{ref} (i)}
\end{equation*}
 where, $S = {k \in [1, M] : Nref (k) \subseteq Nsys(k)}$ and $M$ is the total number of input segments.

\section{Results and discussion}
\label{sec:results}

\begin{table}[h]
\centering
\caption{DCF (in $\%$) for SAD}
\resizebox{0.6\columnwidth}{!}{%
\begin{tabular}{|c|c|c|} \hline
& Dev & Eval\\ \hline
Baseline & 12.5 & 13.6  \\
Proposed & 5.96 & 5.33\\ \hline
\end{tabular}
}
\label{tab:results_sad}
\end{table}

Table~\ref{tab:results_sad} shows the results of the systems submitted for both SAD and SID  tasks to the challenge. The proposed CNN-based SAD system gave a DCF value of $5.96\%$ and $5.33\%$ on development and test data.  These values are $6.5\%$ and $8.27\%$ better than the baseline systems, respectively in terms of absolute number. The result suggests that for supervised VAD approach, even in noisy condition, the neural network-based approach is perhaps better than other modelling approaches. This can be validated by the results of the other submitted systems. The experiments also show that the network architecture and the parameters - size and shift of the chunks are also vital to achieve the best result.

\begin{figure}[h]
    \centering
    \includegraphics[width=\columnwidth]{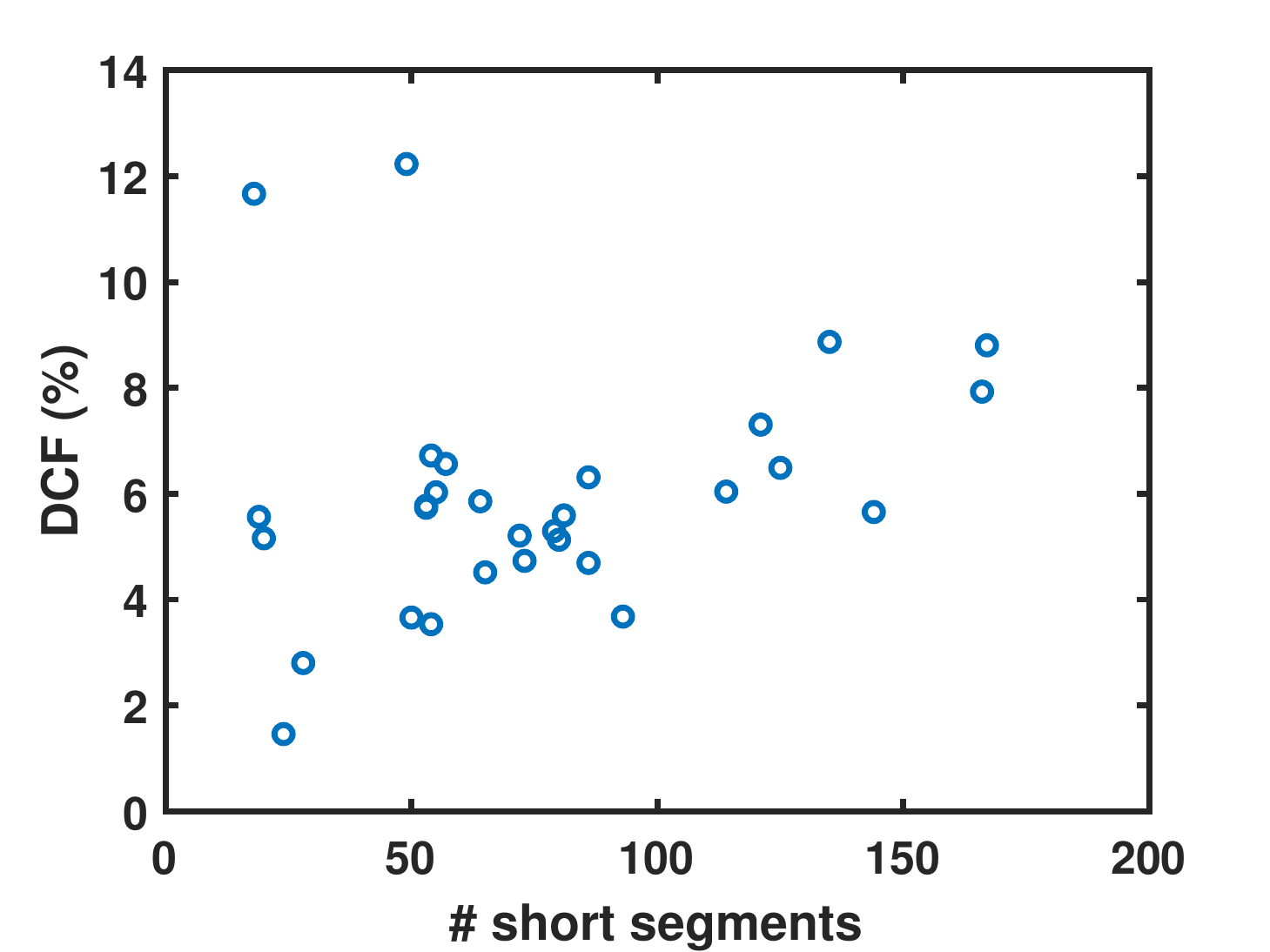}
    \caption{Scatter plot showing the DCF measure as a function of the total number of short segments (speech/non-speech) in each file on development set}
    \label{fig:scatter_sad}
\end{figure}

Figure~\ref{fig:scatter_sad} gives an insight of the DCF for individual files. In the figure, the x-axis shows the total number of short segments ($<0.2s$) for each file on the devlelopment data. Except for the two outliers, a linear relationship is observed between the duration and the DCF values. Hence, it can be inferred that the main factor for difference in DCF among the files is the  presence of short segments. 

On SID task, as shown in Table~\ref{tab:results_sid}, the top-5 accuracy of the submitted system is better with an absolute difference of $6.87\%$ on development data and $9.92\%$ on evaluation data. The decision based on the hypothesization combined voting rule is shown to be improving the performance of the system.
\begin{table}[h]
\centering
\caption{Top-5 detection accuracy (in $\%$) for SID task}
\resizebox{0.6\columnwidth}{!}{%
\begin{tabular}{|c|c|c|} \hline
& Dev & Eval \\ \hline
Baseline & 75.2 & 72.5 \\
Proposed & 82.07 & 82.42\\ \hline
\end{tabular}
}
\label{tab:results_sid}
\end{table}


A detailed analysis is performed using the results obtained from the best system. Though the accuracy seems to be good, it is still computed using the top-5 retrieved results. Results are likely to be worse when the one-best speaker is identified.
\begin{table}[h]
\centering
\caption{Accuracy as a function of top-N retrieval on development data for different N values}
\resizebox{\columnwidth}{!}{%
\begin{tabular}{|c|c|c|c|c|c|} \hline
& Top-5 & Top-4 & Top-3 & Top-2 & Top-1 \\ \hline
Accuracy (\%) & 82.1 & 78.7 & 73.3 & 62.8 & 40.4 \\ \hline
\end{tabular}
}
\label{tab:accuracy}
\end{table}Table~\ref{tab:accuracy} shows the degradation of the accuracy when the N value in the top-N retrieval is decreased. An absolute reduction in the accuracy by approximately $4\%$ is seen when N is reduced from 5 to 4 and from 4 to 3. But the degradation is significant by an absolute value of about $10\%$ and $20\%$ when 2-best and 1-best retrieval are used, respectively. This significant degradation may be attributed to the imbalance in the speaker data used for training.

\begin{figure}[h]
    \centering
    \includegraphics[width=\columnwidth]{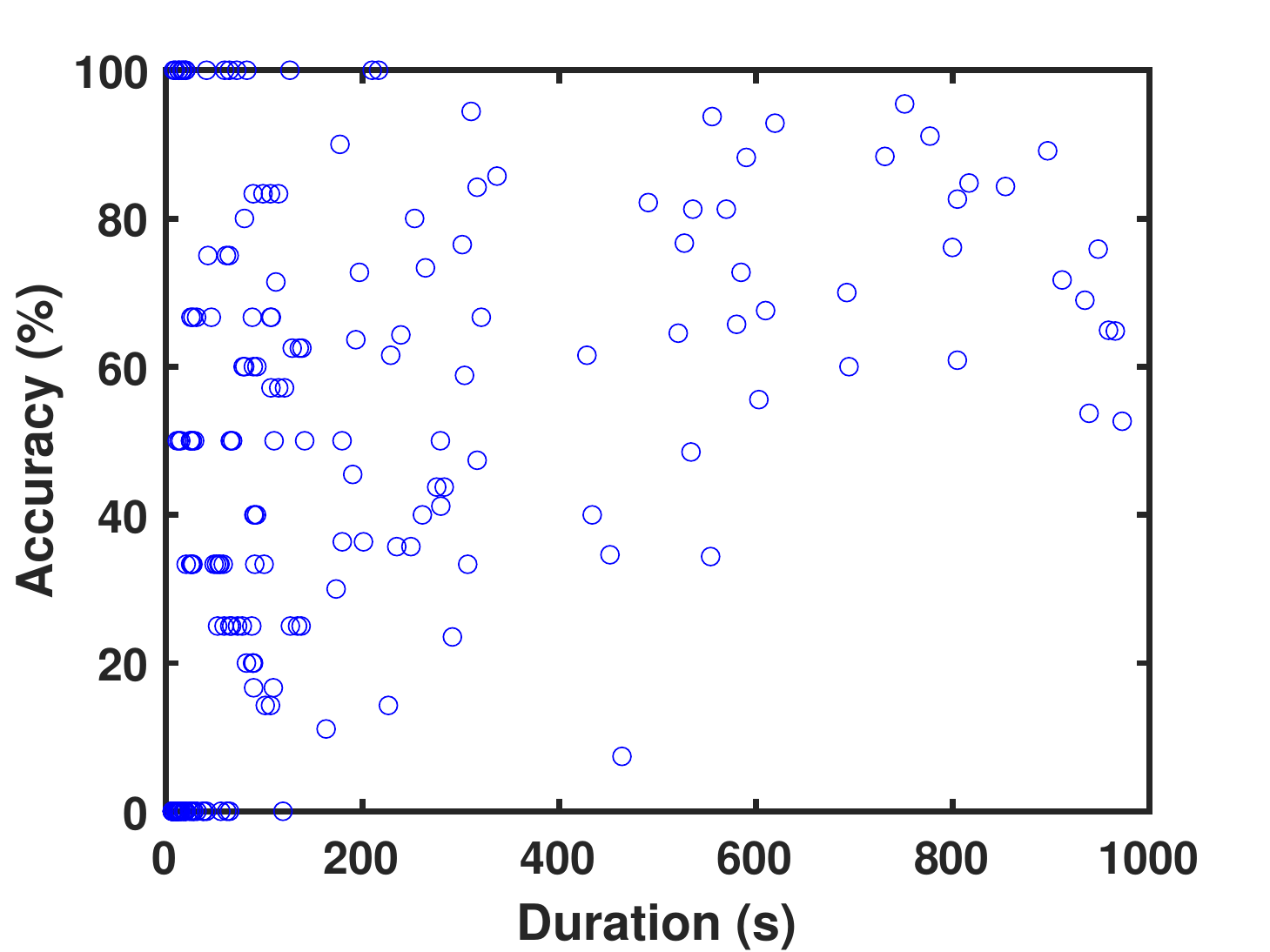}
    \caption{Top-5 accuracy as a function of the duration of the train speakers. Speakers with more than $1000s$ are removed for plotting.}
    \label{fig:scatter}
\end{figure}

Two challenging factors in the dataset are 1. High imbalance in terms of speaker duration in the training data 2. Short utterance for testing. The effects of the two challenges are studied. Figure~\ref{fig:scatter} is a plot of accuracy as a function of the duration of the train speaker. The values in the plot are computed on the development data. The plot shows that the dataset is highly imbalanced in terms of per speaker training duration. Despite this imbalance, about $40\%$ of speakers' data with a training duration of less than $60$ seconds had more than $50\%$ accuracy. For speakers' data more than $500s$, about $80\%$ of files had more than $75\%$ accuracy. 

\begin{figure}[h]
    \centering
    \includegraphics[width=\columnwidth]{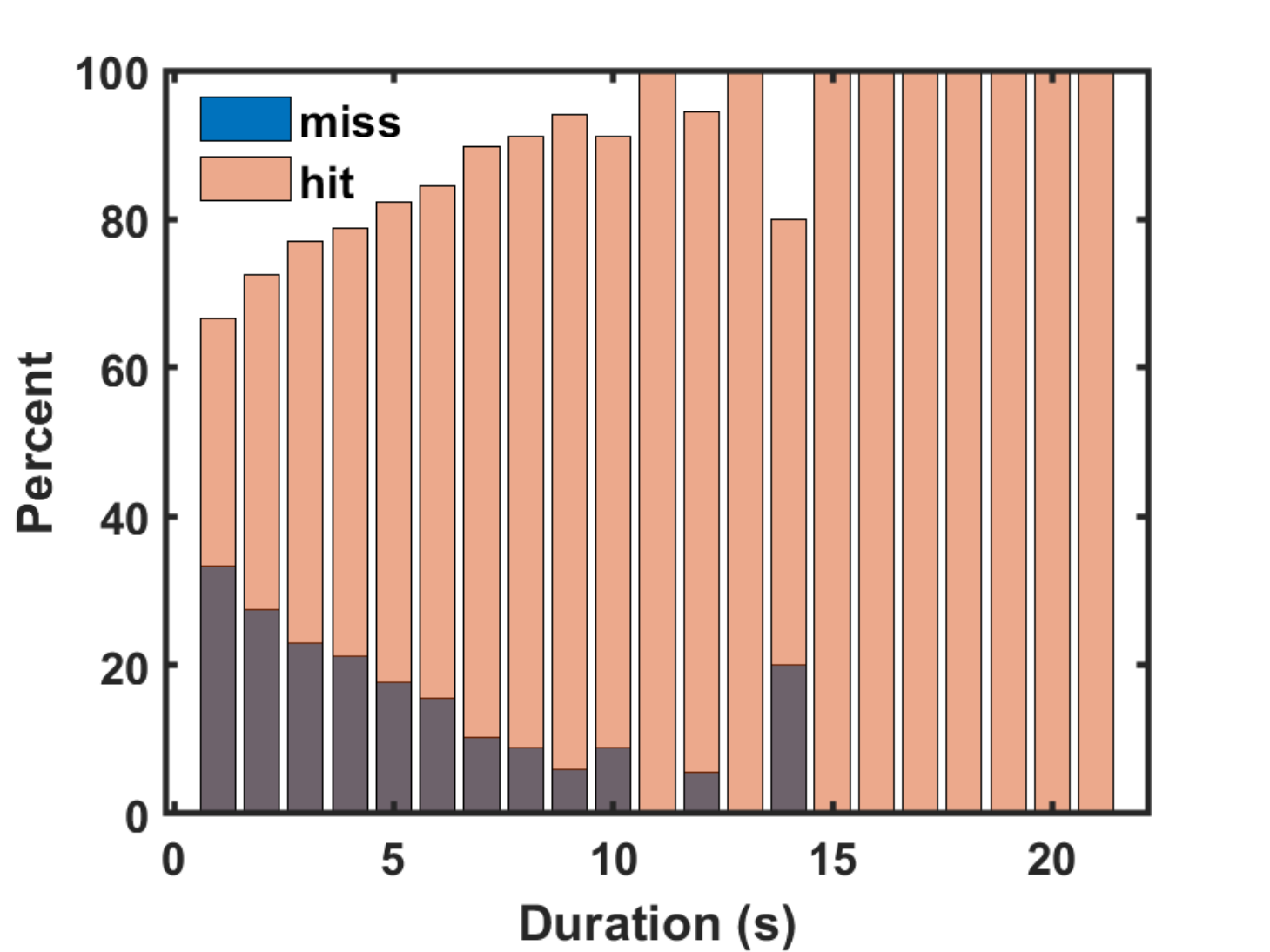}
    \caption{Hit and miss rates as a function of duration of the test utterance}
    \label{fig:bar}
\end{figure}
The second challenging factor is the short duration of the test utterances. The duration of the test utterances ranges from $1s$ to $20s$. It is well known that the speaker identification accuracy degrades when the duration of test utterances is short. Figure~\ref{fig:bar} shows the percentage of hits and misses of files with a set of duration bins. The test files are segregated into a set of bins based on the duration. Hit and miss rates are computed for each bin. It can be seen that when the duration of the test utterance is more than $10s$, the accuracy is close to $100\%$. The rate at which the accuracy degrades as a function of test audio duration is shown in the figure. It is interesting to see that, even in the shortest duration $(2s)$ bin, hit percentage is more than $50\%$.

The proposed system for SAD system differs from existing CNN methods mostly based on the chunking. A CNN-based system was proposed in \cite{Vafeiadis2019} for Fearless Step Phase 1 challenge. With respect to the feature, the work used a longer ($1s$) with a higher spectrogram resolution. Further, they have used an RNN in place of the fully connected layers used in this approach.

Most CNN methods are proposed for speaker verification task. The approaches are either embedding-based using a back-end discriminative training, or the network is a Siamese network \cite{salehghaffari2018speaker}. However, in this work, a two-level voting rule-based method is used which improves the identification accuracy.

\section{Conclusion}
\label{sec:conc}
The results on the SAD and SID tasks suggest that the convolutional neural network is well suited for noisy datasets under supervised training conditions. SAD on chunked spectrogram seem to be a reliable approach to identify speech and non-speech regions. This approach can easily be extended to online systems that can work real-time. On SID task, the proposed two-level voting rule seems to advantageous than score aggregation-based approaches. Despite achieving a top-5 identification accuracy of $82\%$, the accuracy drops to $40\%$ when one-best identification is made. This drop is mainly attributed to a very small amount of training data for few speakers. Nevertheless, the proposed approach is shown to be effective on the challenging Fearless Steps corpus.

\bibliographystyle{IEEEtran}

\bibliography{mybib}

\end{document}